\documentclass[aps,prb,a4paper,10pt,twocolumn,showpacs,floatfix,superscriptaddress,preprintnumbers,longbibliography]{revtex4-2}
\usepackage{float}
\usepackage{dcolumn,graphicx,color,booktabs,microtype,afterpage}
\usepackage{amssymb}
\usepackage{amsmath}
\usepackage[charter,greekuppercase=italicized]{mathdesign}
\usepackage{sidecap}
\usepackage[mathlines]{lineno}

\graphicspath{{./}{figure/}}
\renewcommand{\tablename}{Table}
\makeatletter\renewcommand{\fnum@figure}[1]{\figurename~\thefigure.~}\makeatother
\makeatletter\renewcommand{\fnum@table}[1]{\tablename~\thetable.}\makeatother

\newcount\hh \newcount\mm
\hh=\time \divide\hh by 60
\mm=\hh \multiply\mm by 60 \mm=-\mm
\advance\mm by \time
\def\now{\number\hh:\ifnum\mm<10{}0\fi\number\mm}

\usepackage[colorlinks,plainpages=false,linkcolor=blue,urlcolor=blue,citecolor=blue,pdfpagemode=UseNone,pdfstartview=FitBH]{hyperref}
\usepackage{nicefrac}

\newcommand{\tcr}[1]{\textcolor{black}{#1}}

\begin{document}

\makeatletter\renewcommand{\ps@plain}{%
\def\@evenhead{\hfill\itshape\rightmark}%
\def\@oddhead{\itshape\leftmark\hfill}%
\renewcommand{\@evenfoot}{\hfill\small{--~\thepage~--}\hfill}%
\renewcommand{\@oddfoot}{\hfill\small{--~\thepage~--}\hfill}%
}\makeatother\pagestyle{plain}

\preprint{\textit{Preprint: \today, \now}} 
\title{Evidence of fully-gapped superconductivity in NbReSi: \\ A combined $\mu$SR and NMR study}

%
\author{T.\ Shang}\email[Corresponding authors:\\]{tshang@phy.ecnu.edu.cn}
\affiliation{Key Laboratory of Polar Materials and Devices (MOE), School of Physics and Electronic Science, East China Normal University, Shanghai 200241, China}
%
\author{D. Tay}
\affiliation{Laboratorium f\"ur Festk\"orperphysik, ETH Z\"urich, CH-8093 Z\"urich, Switzerland}
\author{H. Su}
\affiliation{Center for Correlated Matter and Department of Physics, Zhejiang University, Hangzhou 310058, China}
\affiliation{Zhejiang Province Key Laboratory of Quantum Technology and Device, Department of Physics, Zhejiang University, Hangzhou 310058, China}
%
%
%
%
%
%
%
%
%
\author{H. Q. Yuan}
\affiliation{Center for Correlated Matter and Department of Physics, Zhejiang University, Hangzhou 310058, China}
\affiliation{Zhejiang Province Key Laboratory of Quantum Technology and Device, Department of Physics, Zhejiang University, Hangzhou 310058, China}
\affiliation{Collaborative Innovation Center of Advanced Microstructures, Nanjing University, Nanjing, 210093, China}
\affiliation{State Key Laboratory of Silicon Materials, Zhejiang University, Hangzhou 310058, China}

\author{T.\ Shiroka}
\affiliation{Laboratorium f\"ur Festk\"orperphysik, ETH Z\"urich, CH-8093 Z\"urich, Switzerland}
\affiliation{Laboratory for Muon-Spin Spectroscopy, Paul Scherrer Institut, Villigen PSI, Switzerland}
\begin{abstract}
We report a comprehensive study of the noncentrosymmetric NbReSi 
superconductor by means of muon-spin rotation and relaxation ($\mu$SR) 
and nuclear magnetic resonance (NMR) techniques. NbReSi is a 
bulk superconductor with $T_c = 6.5$\,K, characterized by a large 
upper critical field, 
which exceeds the Pauli limit. 
Both the superfluid density $\rho_\mathrm{sc}(T)$ (determined via 
transverse-field $\mu$SR) and the spin-lattice relaxation rate $T_1^{-1}(T)$ 
(determined via NMR) suggest a nodeless superconductivity (SC) in NbReSi. 
We also find signatures of multigap SC, here evidenced 
by the field-dependent muon-spin relaxation rate and the electronic 
specific-heat coefficient.  
The absence of spontaneous magnetic fields below 
$T_c$, as evinced from zero-field $\mu$SR measurements, indicates a preserved time-reversal symmetry in 
the superconducting state of NbReSi. 
Finally, we discuss possible reasons for the unusually large upper 
critical field of NbReSi, most likely arising from its anisotropic 
crystal structure.
\end{abstract}

\maketitle\enlargethispage{3pt}

\vspace{-5pt}
\section{\label{sec:Introduction}Introduction}\enlargethispage{8pt}
Superconductors whose crystal structures lack an inversion center, 
known as noncentrosymmetric superconductors (NCSCs), represent an 
attractive platform for investigating unconventional- and topological 
superconductivity (SC)~\cite{Bauer2012,Smidman2017,Ghosh2020b,Kim2018,Sun2015,Ali2014,Sato2009,Tanaka2010,Sato2017,Qi2011,Kallin2016}. 
Besides SC, noncentrosymmetric materials are among the best 
candidates for studying also topological phenomena. For example, 
Weyl fermions were discovered as quasiparticles in Ta(As,P) and 
Nb(As,P) noncentrosymmetric single crystals~\cite{Xu2015a,Xu2015b,Lv2015,Xu2016,Souma2016}. 
In NCSCs, a lack of inversion center sets the scene for a variety of exotic properties, e.g., 
nodes in the superconducting gap~\cite{yuan2006,nishiyama2007,bonalde2005CePt3Si,Shang2020}, 
multigap SC~\cite{kuroiwa2008}, upper critical fields beyond the Pauli 
limit~\cite{Carnicom2018,Bauer2004,Su2021}, and breaking of time-reversal 
symmetry (TRS) in the superconducting 
state~\cite{Shang2020,Hillier2009,Barker2015,Shang2020b,Singh2014,Shang2018a,Shang2018b}.

In some NCSCs, the above exotic properties are closely related 
to admixtures of spin-singlet and spin-triplet superconducting pairing, 
here tuned by the antisymmetric spin-orbit coupling 
(ASOC)~\cite{Bauer2012,Smidman2017,Ghosh2020b}. In most other cases, 
however, such connection seems very weak and the superconducting 
properties resemble those of conventional superconductors, 
characterized by a fully developed energy gap and a preserved TRS. 
A notable exception is CaPtAs which, below $T_c$, exhibits both TRS 
breaking and superconducting gap nodes~\cite{Shang2020,Xie2019}. 
In general, the causes behind TRS breaking in NCSCs are not yet fully  understood and remain an intriguing 
open question. 

After the discovery of TRS breaking in elementary rhenium and in 
$\alpha$-Mn-type Re$T$ superconductors ($T$ = transition 
metal)~\cite{Singh2014,Shang2018a,Shang2018b}, many other Re-based 
superconductors have been systematically investigated by means of 
muon-spin relaxation and rotation ($\mu$SR).  
Later on, $\mu$SR studies on Re$_{1-x}$Mo$_x$ alloys ($0 \le x \le 1$) 
revealed that TRS is broken only in cases of a high rhenium content ($x < 0.12$),
surprisingly corresponding to simple centrosymmetric structures~\cite{Shang2020ReMo}. 
For $x > 0.12$, instead, all the Re-Mo alloys preserve TRS in their 
superconducting state, independent of the centro- or noncentrosymmetric 
crystal structure~\cite{Shang2020ReMo}. Recently, the centrosymmetric 
Re$_3$B and noncentrosymmetric Re$_7$B$_3$ superconductors were 
systematically studied and shown to exhibit nodeless SC with multiple 
energy gaps~\cite{Shang2021}. 
In both cases, the lack of spontaneous magnetic fields below $T_c$ 
indicates that, unlike in Re$T$ or in elementary rhenium, the TRS is 
preserved. Such a selective occurrence of TRS breaking in Re-based superconductors,  
independent of the noncentrosymmetric structure (and thus of ASOC), is 
puzzling and not yet fully understood, clearly demanding further 
investigations. 

\begin{figure}[!thp]
\centering
    \vspace{-1ex}%
	\includegraphics[width=0.47\textwidth,angle=0]{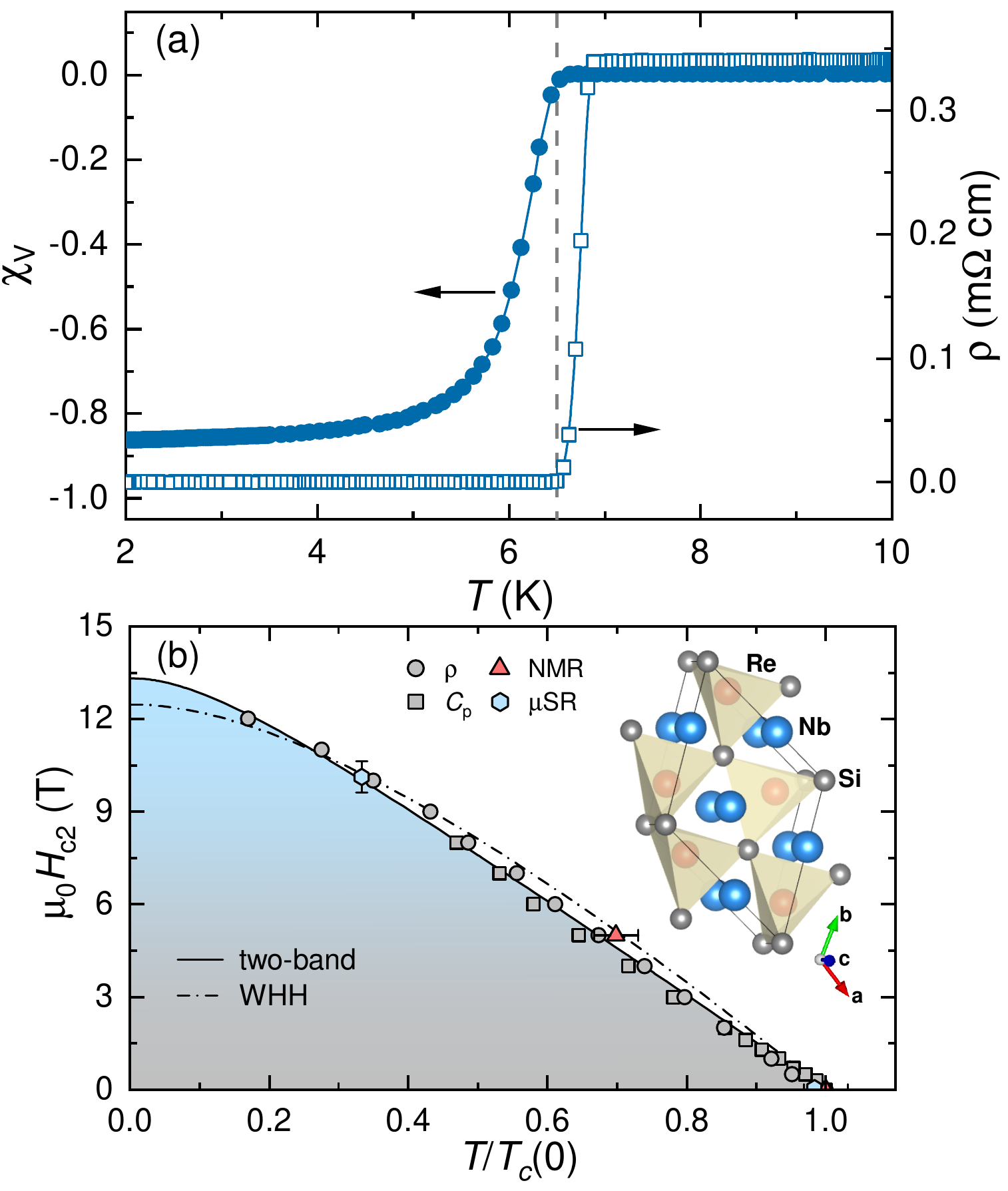}
	\caption{\label{fig:Hc}(a) Temperature-dependent magnetic susceptibility 
	$\chi_\mathrm{V}(T)$ (left-axis) and electrical resistivity $\rho(T)$ (right-axis).
	While $\rho(T)$ was measured in a zero-field condition, $\chi_\mathrm{V}(T)$ data 
	were collected in a magnetic field of 1\,mT, applied after 
	zero-field cooling (ZFC). The dashed line indicates the $T_c = 6.5$\,K. 
    (b) Upper critical field $H_\mathrm{c2}$ vs reduced 
	temperature $T_c$/$T_c(0)$ for NbReSi.
	The solid- and dash-dotted lines represent fits to the two-band- and 
	WHH model.
	The $T_c$ values determined from $\mu$SR- and NMR measurements 
	(this work) are highly consistent with the values determined from 
	electrical-resistivity- and heat-capacity measurements (data 
	taken from Ref.~\onlinecite{Su2021}). The inset shows the crystal
	structure of NbReSi, with the lines marking its unit cell. }
\end{figure}

The NbReSi superconductor represents yet another
candidate material for studying the TRS breaking effect in the 
Re-based family of superconductors. NbReSi crystallizes in a hexagonal 
ZrNiAl-type crystal structure with space group $P\overline{6}2m$ (No.~189) 
[see inset in Fig.~\ref{fig:Hc}(b)]~\cite{Osti1974}. 
Although its SC was reported in 1980s~\cite{Rao1985}, its physical 
properties were systematically studied only recently~\cite{Su2021}.
As shown in Fig.~\ref{fig:Hc}(a), both magnetic-susceptibility and 
electrical-resistivity data indicate a $T_c = 6.5$\,K in NbReSi.
The temperature evolution of the upper critical field $H_\mathrm{c2}(T)$, 
as established by electrical-resistivity- and heat-capacity measurements, 
is reported in Fig.~\ref{fig:Hc}(b), with $H_\mathrm{c2}(0)$ 
being larger than the weak-coupling Pauli limit value 
(i.e., 1.86$k_\mathrm{B} T_c = 12.1$\,T). 
This indicates that the effects of paramagnetic limiting may be either 
reduced or entirely absent in NbReSi and, hence, that it can 
possibly exihibit unconventional superconductivity. 

Although selected properties of NbReSi have been investigated by 
different techniques~\cite{Su2021}, at a microscopic level its 
superconducting properties, in particular, the superconducting order 
parameter, have not been explored. In this paper, we report on an 
extensive study of NbReSi carried out mostly by $\mu$SR and nuclear 
magnetic resonance (NMR) methods, both in its superconducting- and 
normal states. 
Despite a noncentrosymmetric crystal structure, NbReSi is shown to 
be a moderately correlated electron material, which adopts a fully 
gapped superconducting state with preserved TRS.   

%

\section{Experimental details\label{sec:details}}\enlargethispage{8pt}

Polycrystalline NbReSi samples were prepared by the arc-melting method. 
The crystal structure and phase purity were checked by powder 
x-ray diffraction. 
The bulk SC was characterized by electrical-resistivity-, heat-ca\-pac\-i\-\mbox{ty-,} 
and magnetization measurements~\cite{Su2021}.  
The bulk $\mu$SR measurements were carried out at the multipurpose 
surface-muon spectrometer (Dolly) of the Swiss muon source at Paul 
Scherrer Institut, Villigen, Switzerland. 
In this study, we performed three kinds of experiments: transverse-field 
(TF)-, zero-field (ZF)-, and longitudinal-field (LF) $\mu$SR measurements.
As to the former, it allowed us determine the temperature evolution of the 
superfluid density. As to the latter two, we aimed at searching for a 
possible breaking of time-reversal symmetry in the superconducting state 
of NbReSi. 
To exclude the possibility of stray magnetic fields during the ZF-$\mu$SR 
measurements, all the magnets were preliminarily degaussed. 
All the $\mu$SR spectra were collected upon heating and were analyzed 
by means of the \texttt{musrfit} software package~\cite{Suter2012}.

$^{93}$Nb NMR measurements, including line shapes and spin-lattice 
relaxation times, were performed on NbReSi in powder form in 
a magnetic field of 5\,T.  
\begin{figure}[htp]
	\centering
	\includegraphics[width=0.48\textwidth,angle= 0]{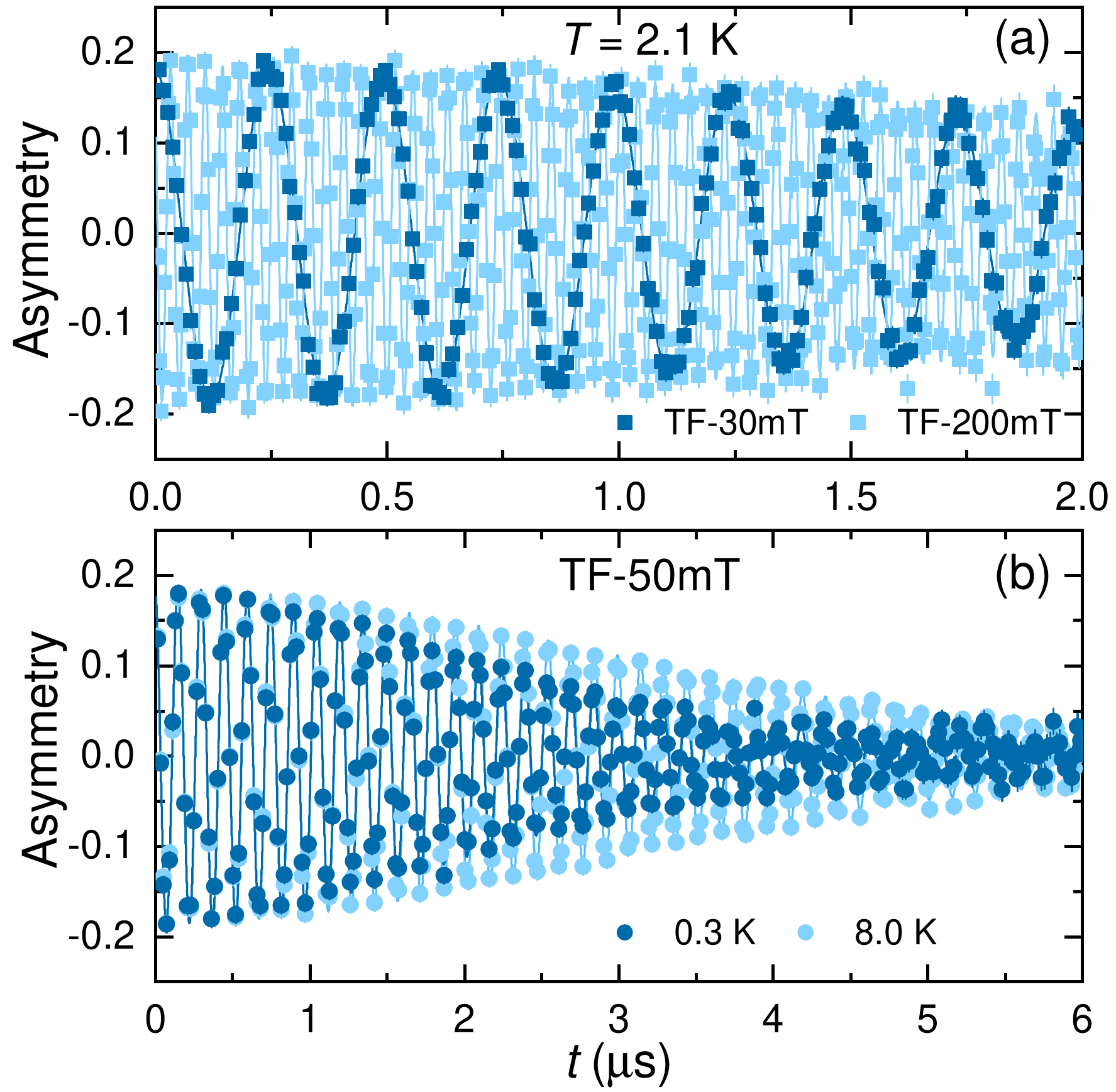}
	\caption{\label{fig:TF-muSR}(a) TF-$\mu$SR spectra of NbReSi 
	collected at $T$ = 2.1\,K (superconducting state) in a field of 
	30 and 200\,mT. (b) TF-$\mu$SR spectra of NbReSi collected 
	in the superconducting- (0.3\,K) and the normal state (8\,K) in 
	an applied magnetic field of 50\,mT. The solid lines 
	represent fits to Eq.~\eqref{eq:TF_muSR}. To clearly show the 
	oscillations at higher fields, the spectra in panel (a) are shown in a 
	time range up to 2\,$\mu$s.}
\end{figure}
To cover the 2 to 300\,K temperature range we used a continuous-flow 
CF-1200 cryostat by Oxford Instruments, with temperatures below 4.2\,K 
being achieved under pumped $^{4}$He conditions. Preliminary resonance 
detuning experiments confirmed the $T_{c}$ of 6.3\,K at 0\,T and of 
4.4\,K at 5\,T. 
The $^{93}$Nb NMR signal was detected by means of a standard spin-echo 
sequence consisting of $\pi/2$ and $\pi$ pulses of 3 and 6\,$\mu$s, 
with recycling delays ranging from 1 to 60\,s in the 2--300\,K
temperature range. 
The lineshapes were obtained via fast Fourier transform (FFT) of 
the echo signal. Spin-lattice relaxation times $T_1$ were measured 
via the inversion-recovery method, using a $\pi$--$\pi/2$--$\pi$ 
pulse sequence. In all the measurements, phase cycling was used to 
systematically minimize the presence of artifacts. 

\section{Results and discussion\label{sec:results}}\enlargethispage{8pt}
\subsection{$\mu$SR study}

%
\begin{figure}[!thp]
	\centering
	\includegraphics[width=0.49\textwidth,angle= 0]{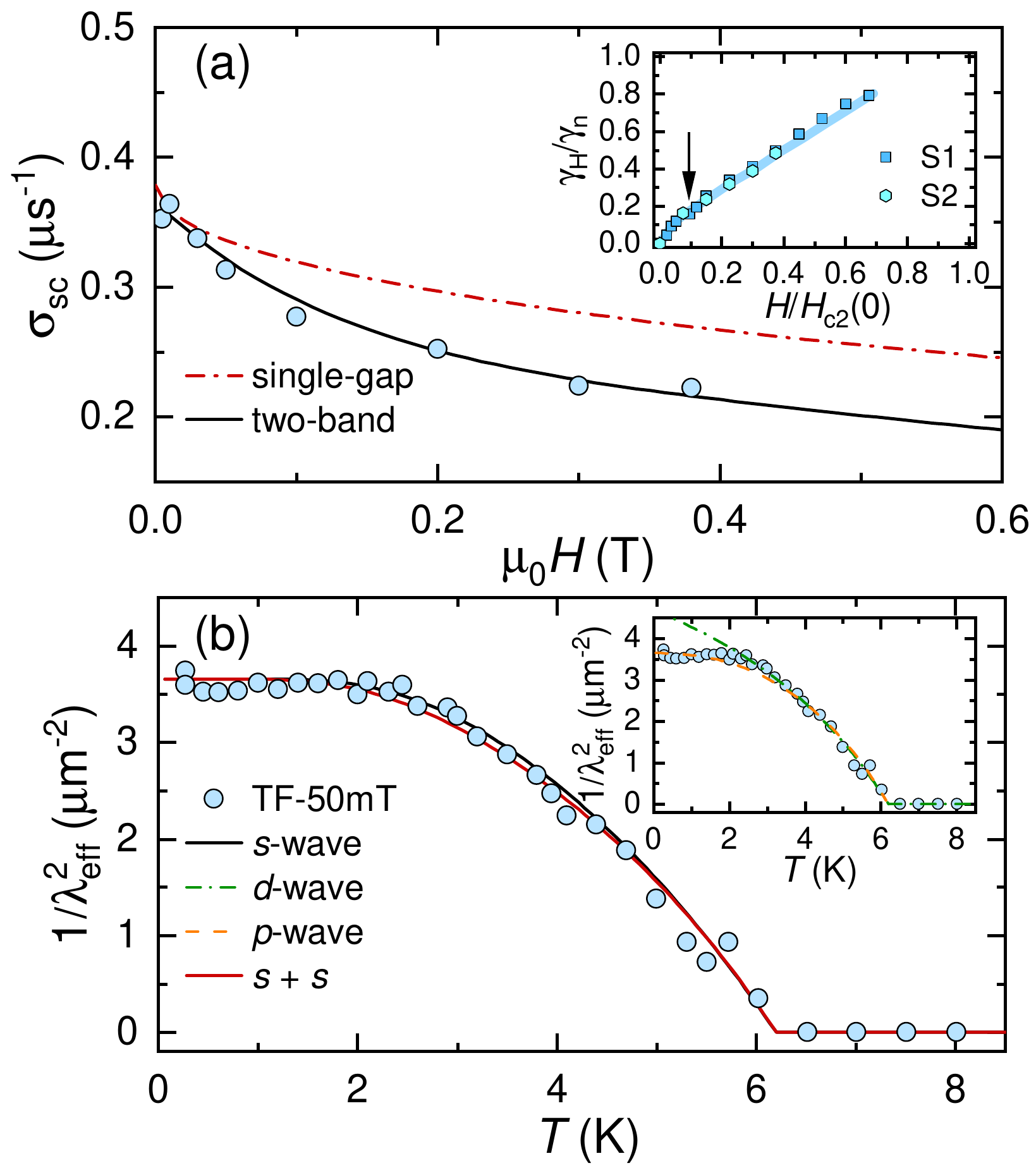}
	\caption{\label{fig:lambda}%
	(a) Field-dependent superconducting Gaussian relaxation rate 
	$\sigma_\mathrm{sc}(H)$. The dash-dotted and solid lines represent 
	fits to a sin\-gle\--gap and a two-band model. The inset plots the 
	normalized specific-heat coefficient $\gamma_\mathrm{H}/\gamma_\mathrm{n}$ 
	vs the reduced magnetic field $H/H_{c2}(0)$ for NbReSi. 
	At a given applied field, $\gamma_\mathrm{H}$ is obtained as the 
	linear extrapolation of electronic specific heat $C_\mathrm{e}/T$ vs 
	$T^2$ in the superconducting state to zero temperature. Data from 
	different samples (denoted as S1 and S2) are highly consistent. 
	The specific heat data can be found in Ref.~\onlinecite{Su2021}. 
	The arrow marks a change of slope at $\mu_0H \sim 1$\,T. 
	(b) Temperature dependence of the NbReSi su\-per\-fluid density. 
	The solid black- and red lines represent fits to a fully-gapped 
	$s$-wave model with one- and two gaps, respectively. 
	The dash-dotted and dashed lines (in the inset) are fits to 
	$p$- and $d$-wave models.}
\end{figure}
%
%

To investigate the superconducting properties of NbReSi at a microscopic 
level, we carried out systematic temperature-dependent $\mu$SR 
measurements in a  
transverse field. 
The optimal field value for such experiments was 
determined via preliminary field-dependent $\mu$SR depolarization-rate 
measurements at 2.1\,K. To track the additional field-distribution 
broadening due to the flux-line lattice (FLL) in the mixed superconducting 
state, a magnetic field (up to 380 mT) was applied in the normal state 
and then the sample was cooled down well below $T_c$,  
where the $\mu$SR spectra were collected. Figure~\ref{fig:TF-muSR}(a) 
shows two representative TF-$\mu$SR spectra collected at 30 and 
200\,mT, in general, modeled by: 
\begin{equation}
	\label{eq:TF_muSR}
	A_\mathrm{TF}(t) =  A_\mathrm{s} e^{- \sigma^2 t^2/2}  \cos(\gamma_{\mu} B_s t + \phi) + A_\mathrm{bg} \cos(\gamma_{\mu} B_\mathrm{bg} t + \phi).
\end{equation}
%

\begin{figure}[t]
	\centering
	\includegraphics[width=0.49\textwidth,angle= 0]{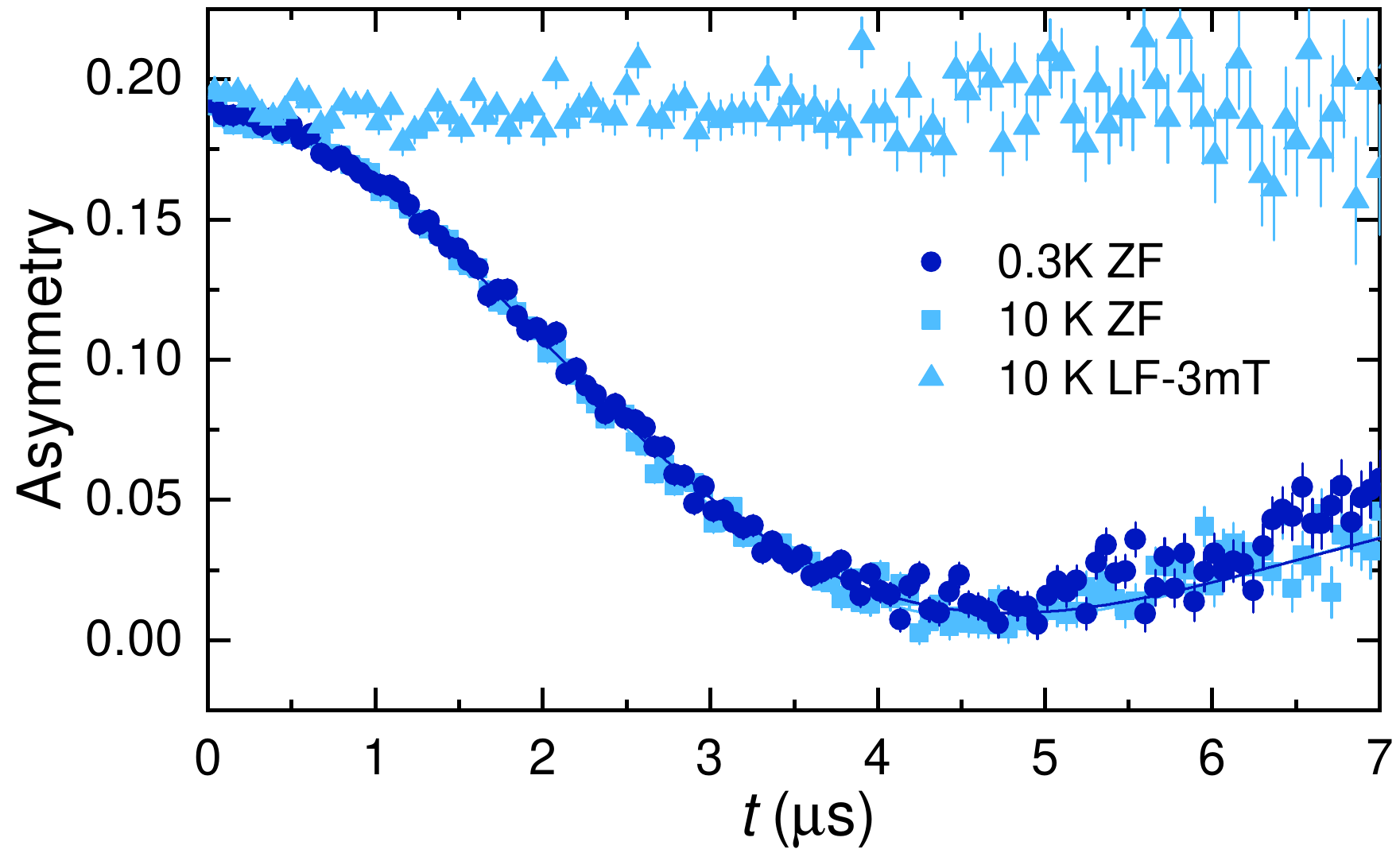}
	\caption{\label{fig:ZF-muSR}ZF-$\mu$SR spectra collected in the superconducting- (0.3\,K) and the normal (10\,K) states for NbReSi. The LF-$\mu$SR in a field of 3\,mT was collected at 0.3\,K after field cooling.}
\end{figure}

Here $A_\mathrm{s}$ (90\%) and $A_\mathrm{bg}$ (10\%) are the sample and 
the background asymmetries, with the latter not undergoing any depolarization. 
$\gamma_{\mu}$/2$\pi$ = 135.53\,MHz/T is the muon gyromagnetic ratio, 
$B_\mathrm{s}$ and $B_\mathrm{bg}$ are the local fields sensed by 
implanted muons in the sample and the sample holder, $\phi$ is a shared 
initial phase, and $\sigma$ is a Gaussian relaxation rate reflecting the 
field distribution inside the sample. In the superconducting state, the 
measured $\sigma$ includes contributions from both the FLL 
($\sigma_\mathrm{sc}$) and a smaller, temperature-independent relaxation, 
due to nuclear moments ($\sigma_\mathrm{n}$) (see also ZF-$\mu$SR below). 
The FLL-related relaxation can be extracted by subtracting the nuclear 
contribution in quadrature, 
$\sigma_\mathrm{sc}$ = $\sqrt{\sigma^{2} - \sigma^{2}_\mathrm{n}}$.

The resulting superconducting Gaussian relaxation rates $\sigma_\mathrm{sc}$ 
vs the applied external magnetic field are summarized in Fig.~\ref{fig:lambda}(a). 
Above the lower critical field $\mu_0H_{c1}$ (10.1\,mT)~\cite{Su2021}, the
relaxation rate decreases continuously. Here, a field of 50\,mT was 
chosen for the temperature-dependent TF-$\mu$SR studies. 
In case of a single-gap superconductor, $\sigma_\mathrm{sc}(H)$ generally 
follows $\sigma_\mathrm{sc} = 0.172 \frac{\gamma_{\mu} \Phi_0}{2\pi}(1-h)[1+1.21(1-\sqrt{h})^3]\lambda^{-2}$~\cite{Barford1988,Brandt2003}, 
where $h = H_\mathrm{appl}/H_{c2}$, with $H_\mathrm{appl}$ being 
the applied magnetic field. As shown by the dash-dotted line in Fig.~\ref{fig:lambda}(a), 
the single-gap model shows a very poor agreement with the experimental data. 
In a two-band model, each band is characterized by its own coherence 
length and a weight $w$ [or ($1-w$)], accounting for the relative contribution of each band 
to the total $\sigma_\mathrm{sc}$ and, hence, to the superfluid 
density~\cite{Serventi2004,Khasanov2014}. 
For such a two-band model, the second moment of the field distribution 
can be calculated  within the framework of the modified London model 
by using the following expression:
	\begin{equation}
		\label{eq:modified_London}
	\langle B^2 \rangle = \frac{\sigma_\mathrm{sc}^2}{\gamma_\mu^2} = B^2 \sum\limits_{q \neq 0} \left[ w \frac{e^{-\frac{q^2\xi_1^2}{2(1-h_1)}}}{1+\frac{q^2\lambda_0^2}{1-h_1}} + (1-w) \frac{-e^{\frac{q^2\xi_2^2}{2(1-h_2)}}}{1+\frac{q^2\lambda_0^2}{1-h_2}} \right]^2.
	\end{equation}
Here, $q = 4\pi/\sqrt{3}a (m\sqrt{3}/2, n + m/2)$ are the reciprocal 
lattice vectors for a hexagonal FLL ($a$ is the inter-vortex distance, 
$m$ and $n$ are integer numbers); $B = \mu_0H$ for $H \gg  H_{c1}$, which 
is the mean field within the FLL; $h_{1(2)} = H/H_{c2,1(2)}$ are the 
reduced fields within band 1(2) (the same as $h$ in the above equation) 
and $\xi_{1(2)}$ are the coherence lengths for the band 1(2).
As shown by the solid line in Fig.~\ref{fig:lambda}(a), with $w = 0.45$, 
the two-band model is in good agreement with the experimental data and 
provides $\lambda_0$ =534(2)\,nm, $\xi_1$ = 18.0(5)\,nm, and 
$\xi_2$ = 5.7(2)\,nm. The derived $\lambda_0$ is comparable 
with the value estimated from the temperature-dependent TF-$\mu$SR 
measurements in Fig.~\ref{fig:lambda}(b). 
The upper critical field of 10.1(5)\,T, calculated from the coherence 
length of the second band, $\mu_0H_{c2} = \Phi_0/(2\pi\xi_2^{2})$, is 
highly consistent 
with the upper critical field determined from bulk 
measurements [see Fig.~\ref{fig:Hc}(b)] 
(The specific-heat data were taken from Ref.~\onlinecite{Su2021}). 
The \emph{virtual} upper critical field $\mu_0H_{c2}^\ast = 1.02(6)$\,T, 
calculated from the coherence length of the first band $\xi_1$, is in 
good agreement with the field value where $\gamma_\mathrm{H}(H)$ shows 
a change in slope [as indicated by an arrow in the inset of 
Fig.~\ref{fig:lambda}(a)]. The virtual 
$H_{c2}^\ast$ corresponds to the critical field which suppresses the small
superconducting gap. Clearly, both the field-dependent $\sigma_\mathrm{sc}(H)$ 
and the electronic specific-heat coefficient $\gamma_\mathrm{H}(H)$ 
suggest the existence of multiple superconducting gaps in NbReSi. 

To further investigate the superconducting pairing in NbReSi, we carried 
out systematic TF-$\mu$SR measurements in an applied field of 50\,mT 
over a wide temperature range. Representative TF-$\mu$SR spectra 
collected in the superconducting (0.3\,K)- and normal (8\,K) state of 
NbReSi are shown in Fig.~\ref{fig:TF-muSR}(b). 
The broadening of the field distribution due to FLL is 
clearly visible in the superconducting state. To quantify it, 
the TF-$\mu$SR spectra were analyzed using again the model given 
by Eq.~\eqref{eq:TF_muSR}.  
In NbReSi, the upper critical field $H_{c2}$ is very large compared 
to the applied TF field (50\,mT). 
Therefore, the effects of overlapping vortex cores with increasing 
field can be ignored when extracting the penetration depth from the 
measured $\sigma_\mathrm{sc}$. 
The effective magnetic penetration depth $\lambda_\mathrm{eff}$ can 
be calculated from $\sigma_\mathrm{sc}^2(T)/\gamma^2_{\mu} = 0.00371\, \Phi_0^2/\lambda^4_{\mathrm{eff}}(T)$~\cite{Barford1988,Brandt2003}.
Figure~\ref{fig:lambda}(b) summarizes the temperature-dependent inverse 
square of the magnetic penetration depth, which is proportional to the 
superfluid density, i.e., $\lambda_\mathrm{eff}^{-2}(T) \propto \rho_\mathrm{sc}(T)$.  
The latter was then analyzed by means of different models, generally described by:
\begin{equation}
	\label{eq:rhos}
	\rho_\mathrm{sc}(T) = 1 + 2\, \Bigg{\langle} \int^{\infty}_{\Delta_\mathrm{k}} \frac{E}{\sqrt{E^2-\Delta_\mathrm{k}^2}} \frac{\partial f}{\partial E} \mathrm{d}E \Bigg{\rangle}_\mathrm{FS}. 
\end{equation}
Here, $f = (1+e^{E/k_\mathrm{B}T})^{-1}$ is the Fermi function and $\langle \rangle_\mathrm{FS}$ represents an average over the Fermi surface~\cite{Tinkham1996}. 
$\Delta_\mathrm{k}(T) = \Delta(T) \delta_\mathrm{k}$ is an angle-dependent gap function, where $\Delta$ is the maximum gap value and $\delta_\mathrm{k}$ is the 
angular dependence of the gap, equal to 1, $\cos2\phi$, and $\sin\theta$ 
for an $s$-, $d$-, and $p$-wave model, respectively, with $\phi$ 
and $\theta$ being the azimuthal angles.
The temperature dependence of the gap is assumed to follow $\Delta(T) = \Delta_0 \mathrm{tanh} \{1.82[1.018(T_\mathrm{c}/T-1)]^{0.51} \}$~\cite{Tinkham1996,Carrington2003}, where $\Delta_0$ is the gap value at 0\,K.

Four different models, including single-gap $s$-, $p$-, and $d$-wave, 
and two-gap $s+s$-wave, were used to describe the $\lambda_\mathrm{eff}^{-2}$$(T)$ 
data. For an $s$- or $p$-wave model, the best fits yield the same 
zero-temperature magnetic penetration depth $\lambda_\mathrm{0} =523(2)$\,nm, 
but different gap values, 1.04(3) and 1.33(5)\,meV, respectively. 
Note that, $\rho_\mathrm{sc}(T)$ is also consistent with a 
dirty-limit model~\cite{Shang2021}, which yields a similar 
superconducting gap [i.e., 0.93(3)\,meV].
For the $d$-wave model, the estimated $\lambda_\mathrm{0}$ and gap value are 460(4)\,nm and 1.28(5)\,meV. 
As can be clearly seen in the inset of Fig.~\ref{fig:lambda}(b), the significant 
deviation of the $p$- or $d$-wave model from the experimental data 
below $\sim 3$\,K and the temperature-independent behavior of $\lambda_\mathrm{eff}^{-2}(T)$ 
for $T < \nicefrac{1}{3}T_c \sim 2$\,K strongly suggest a fully-gapped 
superconductivity in NbReSi.
Since $H_{c2}(T)$, $\sigma_\mathrm{sc}(H)$, and $\gamma_\mathrm{H}(H)$ 
data [see Fig.~\ref{fig:Hc}(b) and Fig.~\ref{fig:lambda}(a)] imply 
the presence of multiple superconducting gaps in NbReSi, 
$\lambda_\mathrm{eff}^{-2}(T)$ was also analyzed using a two-gap 
$s$-wave model. 
In this case, $\rho_\mathrm{sc}(T) = w \rho_\mathrm{sc}^{\Delta_{0,1}}(T) + (1-w) \rho_\mathrm{sc}^{\Delta_{0,2}}(T)$, with $\rho_\mathrm{sc}^{\Delta_{0,1}}$ and 
$\rho_\mathrm{sc}^{\Delta_{0,2}}(T)$ being the superfluid densities related to 
the first- ($\Delta_{0,1}$) and second ($\Delta_{0,2}$) gap, and $w$ a relative weight.
Here, by fixing the weight $w$ = 0.45, as determined from $\sigma_\mathrm{sc}(H)$, the two-gap $s + s$-wave model provides almost identical results to the single-gap $s$-wave model, reflected in two practically overlapping fitting curves in Fig.~\ref{fig:lambda}(b). 
The two-gap model yields $\Delta_{0,1}$ = 0.80(5)\,meV and $\Delta_{0,2}$ = 1.23(5)\,meV. Since the gap sizes are not significantly different ($\Delta_{0,1}$/$\Delta_{0,2}$ $\sim$ 0.7),
this makes it difficult  to discriminate between a single- and a two-gap superconductor based on the temperature-dependent superfluid density alone~\cite{Khasanov2014,Khasanov2020}.  
Nevertheless, as we show above, the two-gap feature in NbReSi is clearly 
reflected in its field-dependent superconducting relaxation rate 
$\sigma_\mathrm{sc}(H)$, specific-heat coefficient $\gamma_\mathrm{H}(H)$, 
and also in the temperature-dependent upper critical field $H_{c2}(T)$. 
The superconducting gap of NbReSi derived from TF-$\mu$SR is similar 
to that of other Re-based superconductors, e.g., Re$T$ 
($T$ = transition metal)~\cite{Shang2018a,Shang2018b,Shang2019,Shang2020ReMo,Shang2021a} 
and rhenium-boron compounds~\cite{Shang2021b}, the latter exhibiting a 
multigap SC, too. 

To search for a possible breaking of the time-reversal symmetry in the 
superconducting state of NbReSi, we compared the ZF-$\mu$SR results in 
the normal- and superconducting states.  
As shown in Fig.~\ref{fig:ZF-muSR}, neither coherent oscillations nor 
fast decays could be identified in the spectra collected above (10\,K) 
and below $T_c$ (0.3\,K), hence implying the lack of any magnetic order 
or fluctuations. 
Normally, in the absence of external fields, the onset of SC does not
imply any changes in the ZF muon-spin relaxation rate. However, if the TRS 
is broken, the onset of spontaneous magnetic fields can be detected 
by ZF-$\mu$SR as an increase in the muon-spin relaxation rate. 
In absence of external fields, the muon-spin relaxation is mainly 
attributed to the randomly oriented nuclear moments, which can be 
modeled by means of a phenomenological 
relaxation function, consisting of a combination of Gaussian- and 
Lorentzian Kubo-Toyabe relaxations~\cite{Kubo1967,Yaouanc2011}, 
$A(t) = A_\mathrm{s}[\frac{1}{3} + \frac{2}{3}(1 - \sigma_\mathrm{ZF}^{2}t^{2} - \Lambda_\mathrm{ZF} t) \mathrm{e}^{(-\frac{\sigma_\mathrm{ZF}^{2}t^{2}}{2} - \Lambda_\mathrm{ZF} t)}] + A_\mathrm{bg}$. 

Here, $A_\mathrm{s}$ and $A_\mathrm{bg}$ are the same as in the 
TF-$\mu$SR case [see Eq.~\eqref{eq:TF_muSR}].
The $\sigma_\mathrm{ZF}$ and $\Lambda_\mathrm{ZF}$ represent the zero-field Gaussian and Lorentzian relaxation rates, respectively.  
As shown by the solid lines in Fig.~\ref{fig:ZF-muSR}, the derived relaxations in the normal- and the superconducting states are almost identical.
This lack of evidence for an additional $\mu$SR relaxation below $T_c$ excludes a possible TRS breaking in the superconducting state of NbReSi.

\subsection{$^{93}$Nb NMR study} 

\begin{figure}[t]
	\centering
	\includegraphics[width=0.50\textwidth,angle=0]{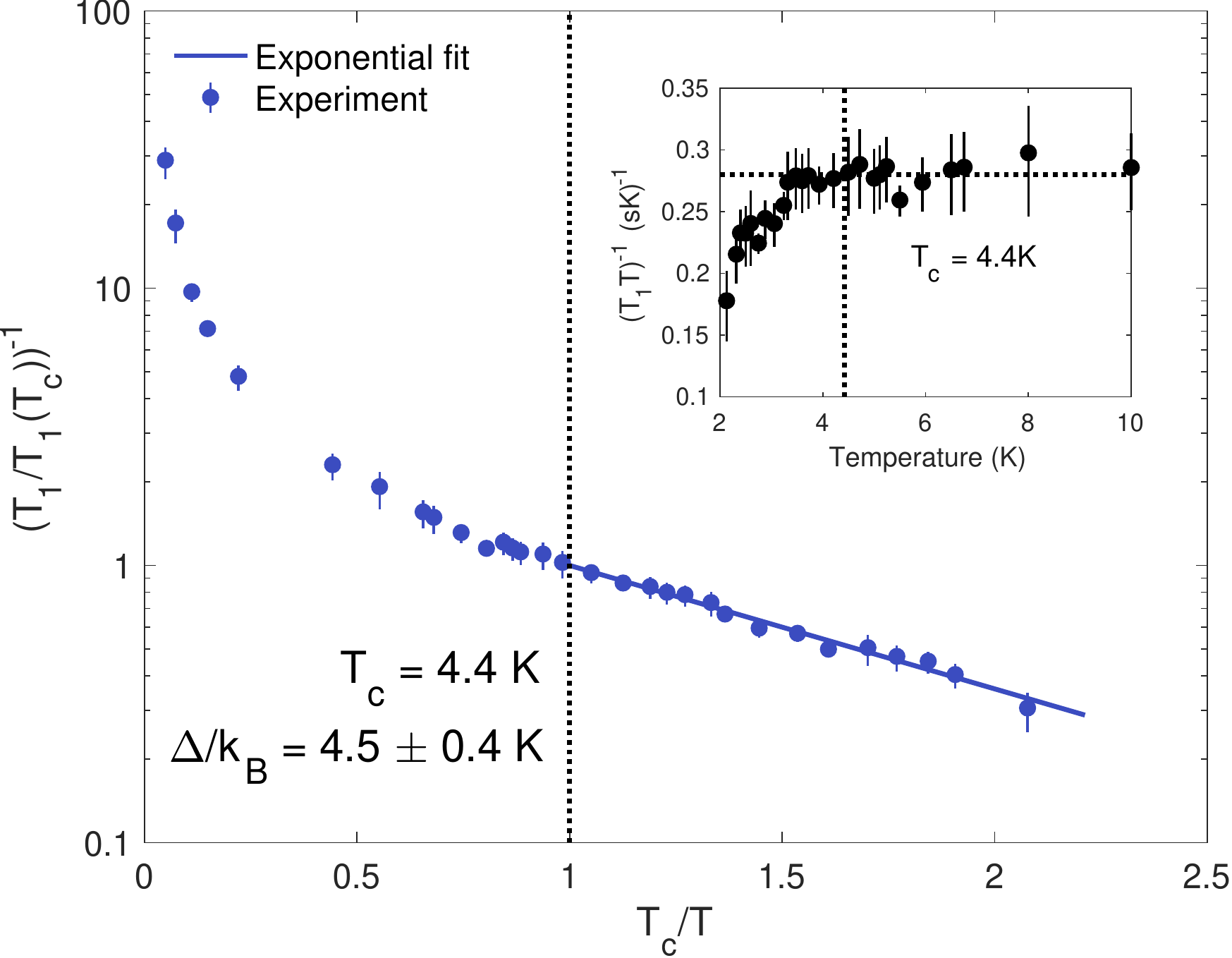}
	\vspace{-2ex}%
	\caption{\label{fig:relaxation}%
		The NMR relaxation rate $T_1^{-1}$ as a function of temperature 
		(here measured at $\mu_0H = 5$\,T) decays by following an 
		exponential law, $T_1^{-1} \propto \exp(-\Delta/k_\mathrm{B}T)$, 
		typical of $s$-wave superconductivity with a fixed gap $\Delta$. 
		The vertical line at 1 refers to the $T_c$ value at 5\,T. 
		Inset: Korringa product $(T_1T)^{-1}$ as a function of temperature. 
		Its constant value above $T_c$ [here, ca.\ $0.28$\,(sK)$^{-1}$] 
		indicates an ideal metallic behavior, followed by a gradual decrease 
		below the superconducting transition. The $T_c$ value 
		determined from $(T_1T)^{-1}$ is consistent those obtained by other 
		techniques [see Fig.~\ref{fig:Hc}(b)]}.
\end{figure}
%

\begin{figure}[t]
	\centering
	\includegraphics[width=0.48\textwidth,angle=0]{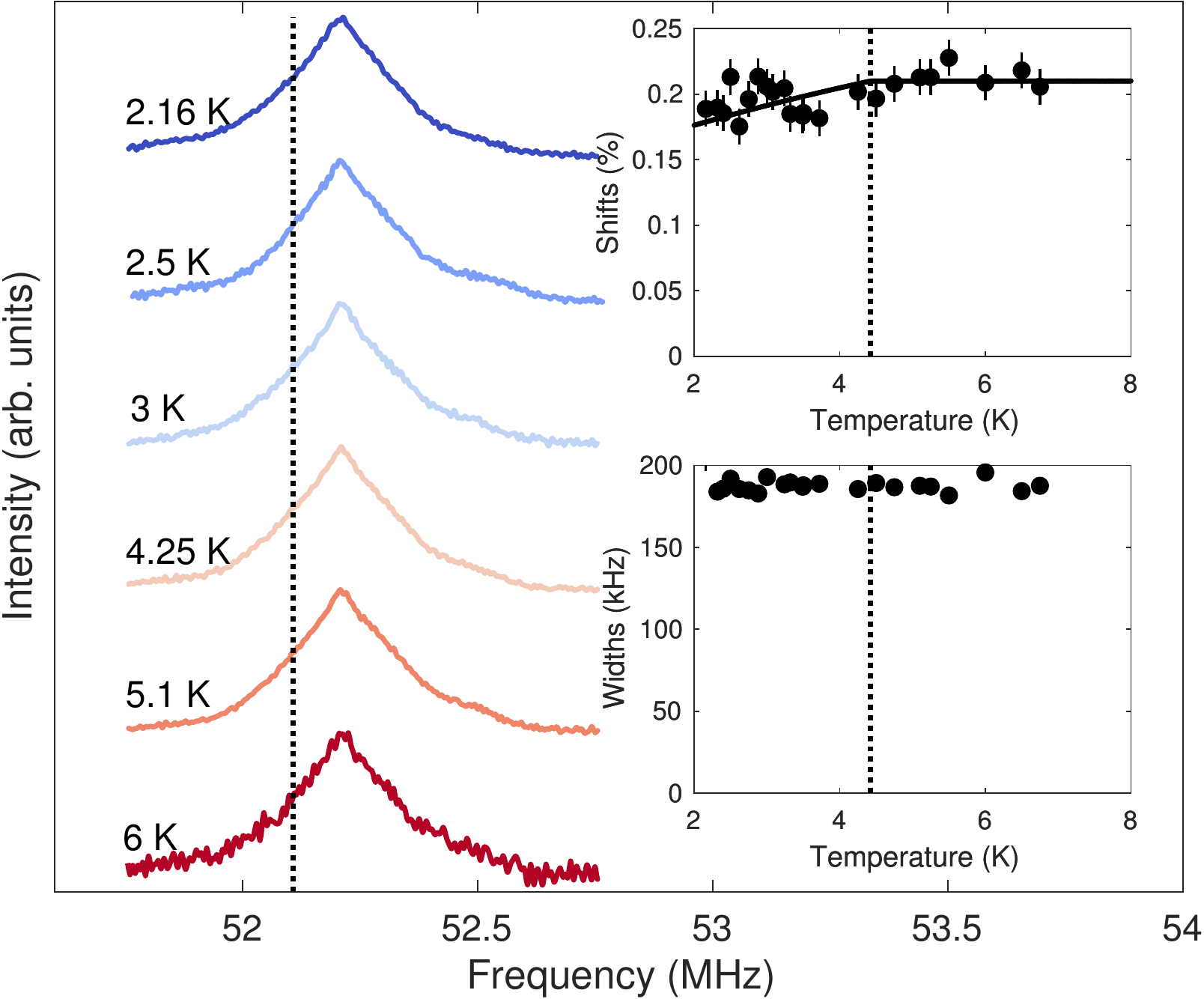}
	\vspace{-2ex}%
	\caption{\label{fig:line}%
		$^{93}$Nb NMR lineshapes collected at various temperatures, covering 
		both the superconducting- and normal states, in a magnetic field of 
		5\,T. The dashed line indicates the reference frequency. The derived 
		NMR Knight shifts and widths vs temperature are shown in the insets. 
		While Knight shifts show a negligible drop below $T_c$, 
		the line widths are independent of temperature.}
\end{figure}

NMR is a versatile technique for investigating the electronic
properties of materials, in particular, their electron correlations,
complementary to $\mu$SR with respect to the probe location, coupling 
to the environment, time window, and field range. 
Here we employed NMR to investigate the normal- and super\-con\-duc\-ting\--sta\-te 
properties of NbReSi, mostly via $^{93}$Nb NMR measurements in a field of 5\,T. 
In selected cases, we conducted also $^{29}$Si NMR ($I = \nicefrac{1}{2}$) 
measurements. 
In either case, the NMR reference frequency $\nu_0$ was determined 
from the $^{27}$Al resonance signal 
in an Al(NO$_3$)$_3$ solution \cite{Harris2002}. 
Successively, the $^{93}$Nb (or $^{29}$Si) NMR shifts were calculated 
with respect to each $\nu_0$ reference frequency. Considering its 
good NMR signal and fast relaxation rate, $^{93}$Nb was used as the 
nucleus of choice for investigating the electronic properties 
of NbReSi.
The NMR line shapes shown in Fig.~\ref{fig:line} 
most likely represent the central transition line and are characterized by 
a relatively large width (ca.\ 200\,kHz). 
No satellite transitions were observed within 1\,MHz on either 
side of the central transition line, indicating that quadrupolar coupling 
is either extremely weak (satellites located within the linewidth of 
the central line) or extremely large (satellites located at $\geq 1$\,MHz 
away from the central transition). Considering the complex 
and asymmetric coordination of Nb atoms, we expect the latter to be the 
case~\cite{Kentgens1997}\footnote{Nb has a $3g$ Wyckoff position, 
with an $m2m$ local symmetry, implying a single $V_{zz}$ component. 
Nb is bonded in a 11-coordinate geometry to six Re and five Si atoms. 
There are two short- (2.87\,\AA) and four long (2.95\,\AA) Nb-Re bonds, 
as well as four short- (2.68\,\AA) and one long (2.78\,\AA) Nb-Si bonds. 
See: https://materialsproject.org/materials/mp-1095061}. 
The central NMR line is sufficient for investigating the superconducting 
properties of NbReSi.

Two main conclusions can be drawn from our NMR study. Firstly, the 
temperature dependence of the Korringa product 
indicates a transition from the metallic to the superconducting 
phase. Secondly, the 
exponential dependence of relaxation rate in the superconducting phase 
confirms the occurrence of a fully-gapped SC phase. 
Our detailed findings are elaborated on below.

\begin{figure}[!thp]
	\centering
	\includegraphics[width=0.4\textwidth,angle=0]{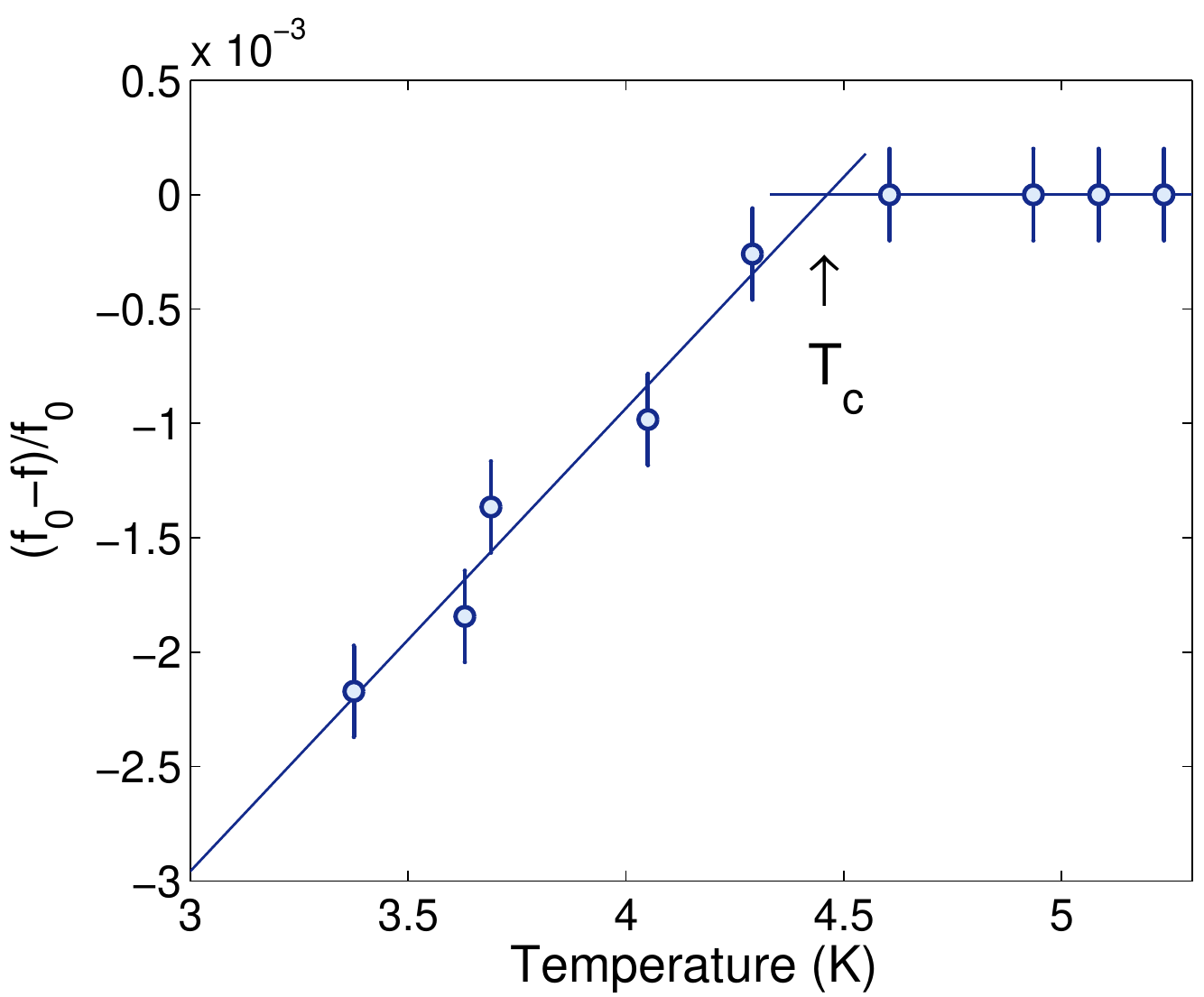}
	\vspace{-2ex}%
	\caption{\label{fig:detuning}%
		\tcr{Frequency detuning of the NMR resonant circuit with temperature. 
			At the onset of superconductivity at $T_c(5\,\mathrm{T}) = 4.4$\,K  
			the magnetic flux is (partially) expelled, equivalent to a lowering 
			of inductance, and hence to an increase in frequency 
			$f = 1/(2\pi \sqrt{LC})$.}}
\end{figure}

\tcr{To determine the $T_{c}(H)$ 
of NbReSi, we use a standard detuning method (see Fig.~\ref{fig:detuning}), 
which gives $T_{c}(5\,\mathrm{T}) = 4.4$\,K. The superconducting transition is 
also confirmed by the Korringa product in the 
$^{93}$Nb case (see inset in Fig.~\ref{fig:relaxation})~\cite{Korringa1950}}. Above $T_c$, 
we observe an ideal behavior [i.e., $(T_1T)^{-1}$ constant], typical of 
standard metals. As the temperature drops below $T_c$ = 4.4\,K, we 
observe a gradual decrease of the $(T_1T)^{-1}$ product, reflecting a 
slowing down of the relaxation rate due to electron pairing, a key 
signature of the superconducting state. 
%

%



%

To determine the nature of the superconducting gap, 
we turn to the relaxation-rate data. 
The relaxation of $^{93}$Nb nuclei can be 
modeled by an exponential function $T_1^{-1} \propto \exp(-\Delta/k_\mathrm{B}T)$ 
(see Fig.~\ref{fig:relaxation}). The fit provides a temperature-independent 
gap $\Delta/k_\mathrm{B} = 4.5 \pm 0.9$\,K (equivalent to 
$\Delta/k_\mathrm{B}T_c = 1.02$), 
lower than that determined by other techniques, most likely 
reflecting an important
quadrupole contribution to relaxation. 
Independent of the gap value,
the exponential decrease of relaxation can be interpreted as 
a critical slowing down of the electronic spin fluctuations. 
The fact that the superconducting gap is constant and temperature 
independent provides strong evidence of conventional $s$-wave 
superconductivity in NbReSi, although NMR is not sensitive enough 
to distinguish single- from multigap SC. 
\tcr{Since unconventional superconductivity would give rise to 
a power-law-dependent relaxation rate vs.\ temperature
(see, e.g., Ref.~\onlinecite{Bauer2014}), 
the exponential decrease of relaxation is sufficient to prove that 
the system possesses a fully-opened gap. 
Nevertheless, for completeness, below we discuss some surprising aspects 
that, although unexpected, can still be reconciled with the theory of 
conventional SC.}

It is surprising that the NMR relaxation data do not exhibit a 
Hebel-Slichter (HS) peak just below the superconducting transition, 
a typical feature of most $s$-wave superconductors.
Several factors might account for
the suppression of the HS peak. 
Firstly, the impurities may play a role in the reduction 
of the HS peak, which might also lead to the weak temperature 
dependence of NMR shifts~\cite{Hines1971}. Secondly, the HS peak can be
smeared out by quadrupole effects, considering that $^{93}$Nb has $I = \nicefrac{9}{2}$.
This smearing 
has been observed in other
noncentrosymmetric superconductors with an $s$-wave gap 
(for instance, W$_3$Al$_2$C shows an HS peak in the 
$I=\nicefrac{1}{2}$ $^{13}$C relaxation, but none in that of 
$I=\nicefrac{5}{2}$ $^{27}$Al~\cite{Tay2021}).
To disentangle the contribution of quadrupole effects to the relaxation 
rate, one has to either measure the same system at lower fields 
or resort to non-quadrupolar nuclei (here, $^{29}$Si). 
The almost quadratic decrease of the signal-to-noise ratio with 
field~\cite{Hoult1976} and the appearance of acoustic ringing at low 
frequencies make this route unpractical. 
On the other hand, numerical calculations suggest that the density of states at Fermi level 
is dominated by the Re-5$d$ and Nb-4$d$ orbitals, the contribution of Si orbitals 
being negligible~\cite{Su2021}. Hence, we expect a much 
weaker magnetic hyperfine coupling to  $^{29}$Si- than to $^{93}$Nb nuclei, 
corresponding to  a much slower relaxation
in the former case.
Indeed, experimentally we find that, in NbReSi, the relaxation rate 
of $^{29}$Si nuclei is prohibitively slow ($\sim 30$\,s at 4\,K, 
becoming exponentially slower below $T_{c}$), thus making the 
use of $^{29}$Si nuclei unfeasible for probing the nuclear relaxation 
in the superconducting state.

\tcr{It is also surprising that both the line shift and width (see 
Fig.~\ref{fig:line}) are almost constant with temperature, with the 
line shapes above and below $T_c$ being virtually indistinguishable. 
The insensitivity of the line shift to the superconducting transition, 
at first seems to suggest a spin-triplet pairing, where electron 
spins maintain their mutual orientation across $T_c$.  However, this
apparent contradiction is resolved by a comparison with
well-known transition-metal superconductors. 
Indeed, as has been shown for tin, vanadium, niobium, etc., their 
total shift is given as the sum of spin- and orbital components 
$K = K_{s} + K_\mathrm{orb}$, with $K_\mathrm{orb}$ being 
the dominant contribution~\cite{Clogston1964}. 
As the latter is unaffected by the electron-spin pairing in the 
superconducting state, this accounts for an almost constant NMR shift 
across $T_c$, yet compatible with a standard $s$-wave pairing. Indeed, 
as shown in Fig.~\ref{fig:line}, we observe a negligible 
decrease in frequency, which could indicate that $K_{s}\ll K_\mathrm{orb}$.  
Other mechanisms, such as disorder, could account for the almost constant NMR shift across $T_c$.
Indeed, in strong\-ly\--cou\-pled systems, 
even tiny amounts of impurities can lead to temperature-independent 
shifts~\cite{Hines1971}. 
Last but not least, sample-heating effects caused by eddy currents 
might also play a role in reducing the observed line shift~\cite{Pustogow2019}.
However, such effects are normally relevant only at very low temperatures in 
good conducting samples (i.e., able to sustain eddy currents), or for 
fast pulse-repetition rates. Since neither the sample nor the measurement 
conditions fully satisfy such requirements, it is unlikely that 
sample heating might explain the weak Knight shifts in NbReSi.}
In general, although some questions about the role
of quadrupole interactions and disorder in suppressing the HS peak 
remain open, 
the exponential dependence of the relaxation rate below $T_c$ 
unambiguously proves the existence of fully-gapped superconductivity in NbReSi.

\subsection{Discussion}
 
According to TF-$\mu$SR measurements at various temperatures, the 
superfluid density $\rho_\mathrm{sc}(T)$ shows an almost 
tem\-per\-a\-ture\--independent behavior below \nicefrac{1}{3}$T_c$ 
[see Fig.~\ref{fig:lambda}(b)], indicating the absence of low-energy 
excitations and thus, a nodeless SC in NbReSi. Both the single-gap $s$- 
and two-gap $s+s$-wave models describe the $\rho_\mathrm{sc}(T)$ data 
very well. However, the field-dependent superconducting Gaussian 
relaxation rate $\sigma_\mathrm{sc}(H)$ and the electronic specific-heat 
coefficient $\gamma_\mathrm{H}(H)$ [see Fig.~\ref{fig:lambda}(a)] 
provide clear evidence of multigap SC in NbReSi, both datasets 
showing a distinct field response compared to a single-gap 
superconductor~\cite{Shang2019c,Shang2020MoPB,Shang2021}.
As indicated by the arrow in the inset of Fig.~\ref{fig:lambda}(a), 
$\gamma_\mathrm{H}(H)$ exhibits a clear change in slope when the applied 
magnetic field (larger than 1\,T) suppresses the small gap, a feature 
recognized as the fingerprint of multigap superconductors. Conversely, 
$\gamma_\mathrm{H}(H)$ would be
mostly linear in the single-gap case. Moreover, 
a single-gap model cannot describe the $\sigma_\mathrm{sc}(H)$ data 
[see main panel of Fig.~\ref{fig:lambda}(a)]. 
At the same time, the two-band model yields an upper critical field 
$\mu_0H_{c2}$(2.1\,K) = 10.1\,T, consistent with the value determined 
from other techniques. The derived \emph{virtual} upper critical field   
$\mu_0H_{c2}^\ast$ = 1.02\,T is in good agreement with the field value 
where $\gamma_\mathrm{H}(H)$ shows a change in slope. 
Here, $H_{c2}^\ast$ corresponds to the critical field which suppresses 
the small superconducting gap. The multigap SC of NbReSi can be further 
inferred from the temperature-dependent upper critical field $H_{c2}(T)$. 
As shown
in Fig.~\ref{fig:Hc}(b), the two-band model~\cite{Gurevich2011} 
is clearly superior to 
the Werthamer-Helfand-Hohenberg (WHH) model~\cite{Werthamer1966} 
over the whole temperature range. 
The analysis of $H_{c2}(T)$ with the two-band model indicates that the 
intra-band and inter-band couplings are $\lambda_{11}$ $\sim$ $\lambda_{22} = 0.17$ 
and $\lambda_{12} = 0.1$, respectively. 
As the inter-band coupling is not much different from intra-band coupling, 
this makes the SC gaps belonging to different electronic bands 
less distinguishable~\cite{Kogan2009}. In addition, as has been found in 
other multigap superconductors~\cite{Shang2019c,Khasanov2014,Shang2020MoPB,Shang2021}, 
a relatively small weight of the second gap, or gap sizes not 
significantly different, make it difficult to discriminate between a 
single- and a two-gap superconductor from the temperature-dependent 
superconducting properties only. 
However, in the NbReSi case, also the electronic band-structure 
calculations support a multigap SC, since they indicate that more than 
four bands cross the Fermi level~\cite{Su2021}.

The two-band model leads to an upper critical field $\mu_0$$H_{c2}(0) = 13.3$\,T, 
which is beyond the weak-coupling Pauli value, i.e., $1.86 k_\mathrm{B}T_c = 12.1$\,T. 
In NCSCs, the antisymmetric spin-orbit coupling allows for the 
occurrence of an admixture of singlet and triplet pairings, which in 
turn can enhance the upper critical field. Consequently, in this case, 
a violation of the Pauli limit hints at the presence of unconventional SC. 
Although in NbReSi the band splitting near the Fermi level is relatively 
large compared to other NCSCs (i.e., $E_\mathrm{ASOC} \sim 180$\,meV)~\cite{Bauer2012,Smidman2017,Su2021}, 
its superconducting pairing is more consistent with spin-singlet, here 
reflected in a fully-gapped superconducting state. Indeed, the  
temperature-dependent zero-field electronic specific heat 
$C_\mathrm{e}(T)/T$, superfluid density $\rho_\mathrm{sc}(T)$, and NMR 
spin relaxation rate $T_1^{-1}(T)$ all suggest a nodeless SC in NbReSi. 
Furthermore, the preserved TRS below $T_c$, as revealed by ZF-$\mu$SR, 
\tcr{suggests the absence of spin-triplet pairing in NbReSi.} 
Therefore, the enhanced $H_{c2}$ of 
NbReSi is unlikely to be caused by a mixed-type of pairing. Strong 
electron correlations can also lead to a large $ H_{c2}$, e.g., the 
noncentrosymmetric CePt$_3$Si and Ce(Rh,Ir)Si$_3$ exhibit $H_{c2}$ 
values far beyond the Pauli limit~\cite{Bauer2004,Kimura2007,Sugitani2006}. 
In NbReSi, however, both the temperature-independent Korringa product 
$(T_1T)^{-1}$ in the normal state and the small electronic specific-heat 
coefficient ($\gamma_\mathrm{n} \sim 8.23$\,mJ/mol-K$^2$~\cite{Su2021}) 
suggest weak electron correlations. A large superconducting gap value 
or a strong electron-phonon coupling may also enhance 
$H_{c2}$~\cite{Bauer2012,Okuda1980}. 
However, the estimated electron-phonon coupling 
$\lambda_\mathrm{ep}$ = 0.66 is rather weak in NbReSi~\cite{Su2021}, 
while its gap value $\Delta_0 =  1.95 k_\mathrm{B}T_c$, as determined 
from TF-$\mu$SR [see Fig.~\ref{fig:lambda}(b)], is not much larger 
than the BCS weak-coupling value (i.e., $1.76 k_\mathrm{B}T_c$). 

Having excluded some common causes of a large $H_{c2}$, 
we examine now the role of anisotropy, considering that a highly 
anisotropic structure can often lead to a sizeable
$H_{c2}$. For instance, in the quasi-one-dimensional Cr-based 
$A_2$Cr$_3$As$_3$ ($A$ = K, Rb, and Cs) superconductors, despite $T_c$s 
in the 2 to 6\,K range, upper critical fields up to 40\,T have been 
reported~\cite{Balakirev2015,Bao2015,Zuo2017,Tang2017,Tang2015}. 
Although triplet pairing was proposed in $A_2$Cr$_3$As$_3$~\cite{Zhi2015,Luo2019,Yang2021}, 
a violation of the Pauli limit is also possible for 
singlet SC. In this case, the spins of Cooper pairs are aligned predominantly 
along the Cr chains, which 
play the role of the
easy magnetization axis, but cause no magnetic order in the normal paramagnetic state. 
This scenario is consistent with the presence of Pauli-limiting pair 
breaking and the strong $H_{c2}(T)$ anisotropy observed in 
K$_2$Cr$_3$As$_3$~\cite{Balakirev2015}.
NbReSi adopts a hexagonal crystal structure ($P\overline{6}2m$, No.~189, 
with in-plane and out-of-plane lattice parameters $a = 6.7194$\,\AA\ and 
$c = 3.4850$\,\AA), which is very similar to the crystal structure of 
$A_2$Cr$_3$As$_3$ ($P\overline{6}m2$, No.~187). Both structures lack an 
inversion center and have a $D_{3h}$ point group. Therefore, the large 
$H_{c2}$ of NbReSi 
is most likely related to its anisotropic crystal structure. 
Such scenario is indirectly supported by the fact that the sister 
compound TaReSi, which also adopts 
a noncentrosymmetric
crystal structure (space group $Ima2$, No.~46), 
but is less anisotropic~\cite{Sajilesh2021}, exhibits a relatively 
small upper critical field, $\mu_0H_{c2}(0)$ = 6.6\,T. To clarify the 
role of anisotropy in enhancing $H_{c2}$, measurements on NbReSi single 
crystals are highly desirable.
Besides the above intrinsic effects, also extrinsic effects may 
enhance the $H_\mathrm{c2}$ of NbReSi. 
For instance, as previously reported in MgB$_2$, single crystals exhibit 
a $H_\mathrm{c2}(0)$ up to 18\,T~\cite{Angst2002}, while a sizable 
impurity scattering introduced by disorder significantly enhances this 
value up to almost 50\,T~\cite{Gurevich2003}. 
To exclude (or confirm) this possibility, again, measurements on 
high-quality single crystals are required. 
Although to date NbReSi single crystals are not available, future work 
on crystal growth could make them accessible.

\section{\label{ssec:Sum}Conclusion}\enlargethispage{8pt}
To summarize, we studied the superconducting properties of the 
noncentrosymmetric NbReSi superconductor by means of the $\mu$SR and 
NMR techniques. The superconducting state of NbReSi is characterized by 
$T_c = 6.5$\,K and an upper critical field $\mu_0H_{c2}(0) = 13.3$\,T.  
The temperature-dependent superfluid density and the NMR 
spin-lattice relaxation 
rate reveal a \emph{nodeless} superconductivity, 
well described by an \emph{isotropic $s$-wave} model. 
\tcr{The NMR Knight shift exhibits only a negligible drop below $T_c$, most 
likely due to a dominant orbital contribution.} 
Field-dependent measurements, including muon-spin relaxation and 
electronic specific-heat coefficient, imply the presence of multiple 
superconducting gaps in NbReSi. This is also supported by the 
temperature dependence of the upper critical field $H_{c2}(T)$. 
The lack of spontaneous magnetic fields below $T_c$ indicates that, 
unlike in Re$T$ or elementary rhenium superconductors, time-reversal 
symmetry is \emph{preserved} in the superconducting state of NbReSi.
In general, the $\mu$SR and the NMR relaxation-rate results are 
highly consistent with a spin-singlet pairing in NbReSi. 
Finally, the violation of Pauli limit in NbReSi is most likely related 
to its anisotropic crystal structure rather than to an unconventional 
type of pairing.

\vspace{1pt}
\begin{acknowledgments}
This work was supported from the Natural Science Foundation of 
Shanghai (Grant Nos.\ 21ZR1420500 and 21JC1402300) and the Schweizerische 
Nationalfonds zur F\"{o}rderung der Wis\-sen\-schaft\-lichen For\-schung 
(SNF) (Grant Nos.\ 200021\_188706 and 206021\_139082). 
H.Q.Y. acknowledge support from the National Key R\&D Program of China (No. 2017YFA0303100 and No. 2016YFA0300202), the Key R\&D Program of Zhejiang Province, China (No. 2021C01002), the National Natural Science Foundation of China (No. 11974306). We acknowledge the allocation of beam time at the Swiss muon source, and thank the scientists of Dolly $\mu$SR spectrometer for their support.
\end{acknowledgments}


\bibliography{NbReSi.bib}

\end{document}